# Chatter reduction in sliding mode tuner controller using skipping surface

K. Fong, R. Leewe and Q.W. Zheng, TRIUMF, V6T2A3, Vancouver, Canada

*Abstract*

TRIUMF ISAC 1 tuning controllers operate using minimum seeking sliding mode controller to minimize the reflected power in their cavities[1,2,3]. As with all extremum seeking algorithms, chatter present in the controller can degrade its performance and cause unnecessary mechanical wear. Various methods has been proposed to reduce this chatter.[4,5] We propose a method which is similar to the "boundary layer method"[4], by observing the rate at which the minimizing function approaches the sliding surface, it is possible to determine whether a change in direction is necessary, thereby reducing the amount of chatter throughout the minimum seeking process.

## *Minimum Seeking Reflected power with Sliding Mode*

The original sliding mode equation is given as[1,2]

$$\dot{\theta} = k_0 \, \text{sgn}\left[\sin\left(\frac{\pi s}{\varepsilon}\right)\right] F(\theta) \quad (1)$$

With $s(t) = F(\theta) + \rho t$  (2)

And the function to be minimized is $F(\theta) = P_r$, the reflected power, whereas $\theta$ is the distance to minimum.

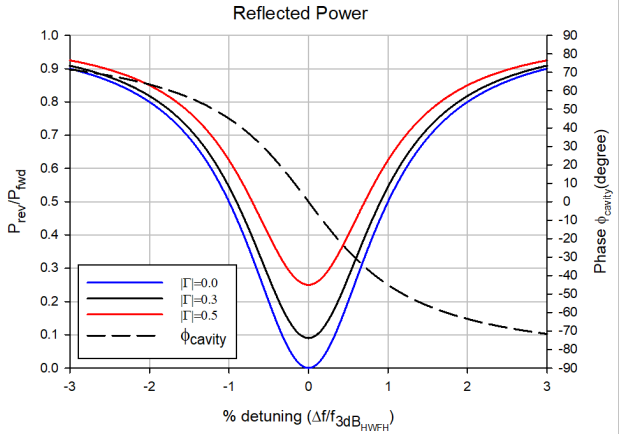

Figure 1. Response of Reflected power and Phase near resonance

It can be seen that $P_r$ is nonlinear and its slope depends on both the amount of detuning and the mismatching. The complicated form of Eq. 1 is useful for theoretical consideration, but in terms of understanding and digital manipulation it is simpler to redefine the mode surfaces are located at $0, \pm 1, \pm 2, ...$. The mode boundaries folded between $-1 < s < 0$ and $0 < s < 1$ by using the C code

*while (s > 1) { s -= 2;}*
*while (s < -1) { s += 2;}*

The sliding function in Eq. 1 becomes

$$s(t) = \frac{1}{\varepsilon}\left[P_r - P_r(0) + \rho t\right] \pm 0.5 \quad (3)$$

Where we have chosen the initial condition $s(0)$ to be $s(0) = \pm 0.5$ This makes $s$ sit in the middle of the two switching surface. The $+\text{ve}$ or the $-\text{ve}$ sign in s(0), which determine the initial tuner moving direction, can be chosen by the initial phase error. $\varepsilon$ is the separation between 2 mode boundaries. The adjustable parameters are $\rho, \varepsilon$, which we shall determine in the following paragraph.

## *Chatter Reduction Parameter Selection: $\rho$*

$s(t)$ stops changing when

$$\frac{d}{dt}P_r = -\rho \quad (4)$$

The finite difference equation is

$$\rho = -\frac{\Delta P_r}{\Delta t} \quad (5)$$

Which can be measured by measuring the difference in $P_r$ after moving the tuner by an amount of $\Delta t$ using the following pseudo code:

```
float r1 = getReflectedPower();
EnableMotot();
float timeFactor = 2.0;
incrPos(slidingMode.gain*deltaTime*timeFactor);
Sleep(1500);    //wait a long while before taking reading
float ck = fabs(r1 - getReflectedPower()) / deltaTime /timeFactor;
incrPos(-slidingMode.gain*deltaTime*2.0);
 return ck;
```

## *Chatter Reduction*

In some instances, at near the switching surfaces, a sort of negative feedback can occurs on *s*, the direction of the tuner can switch rapidly, causing a lot of chatter but little overall movement. For example, when the switching function *s* is close to zero, i.e. from Eq. 4 and Eq. 5,

$s = P_r + \rho t \approx > 0$, $\frac{dP_r}{dt} + \rho \approx < 0$ and $\frac{dP_r}{dt} < 0$, $s$ will drift slowly toward the switch surface $s = 0$. But as soon as $s < 0$, the tuner moves in an opposite direction, causing $\frac{dP_r}{dt} > 0$ and $s$ will arise sharply and back to $s > 0$ and therefore $\frac{dP_r}{dt} < 0$. This process repeats for a while resulting in chatter as illustrated as the green curve in Fig. 2a.

To prevent this from happening, we can add $t = t + \Delta t$ to increase $s$ when $s < 0.1$. We call this "switching surface skipping", as it behaves very much like a stone skipping on the surface of water. The following code does this function to every switching function:

```
if (s>-0.1 && s<0) gtime -= 5*deltaTime;
if (s<0.1 && s>0) gtime += 5*deltaTime;
if (s>0.9)  gtime -= 5*deltaTime;
if (s<-0.9) gtime += 5*deltaTime;
```

which will set back the value of *s* to about 1 sec. as before to eliminate the chatters. Fig. 2b show the improvement in convergence when "switching surface skipping" is used. The blue curves show the reflected power, the green curves show the switching function s for an ordinary, and the red curve show the tuner drive. In both graphs, initially the tuner is moving in the wrong direction. This result in the switching function s rises rapidly until at time=4 when it approaches the switching surface *s*=0. In both cases, due to the choice of the initial reflected power and $\rho$, the switching function s has a tendency to drift downward. In a non de-chattered system, we can see s hovered near 0, resulting in a lot of chatters in the drive and a slower convergence. On the other hand, in the de-chattering "switching surface skip" system, as *s* approaches *s*=0, the de-chattering algorithm pushes the switching function away from the surface. The red curve (drive) switches direction only once and stay in the same direction afterwards. The de-chattering algorithm pushed the slow drift back to the centre between 2 sliding surface. This also result in a faster convergence as the tuner does not spend time switching direction. Figure 3 show the behavior of a de-chattering system when the initial detuning is on the other direction. Since the tuner is moving in the correct direction, the switch function *s* does move up rapidly at t=0 but instead drifting down as in the previous two cases when t > 4. As it approaches the switching surface *s*=-1, the same de-chattering algorithm pushes s back away from the switching surface.

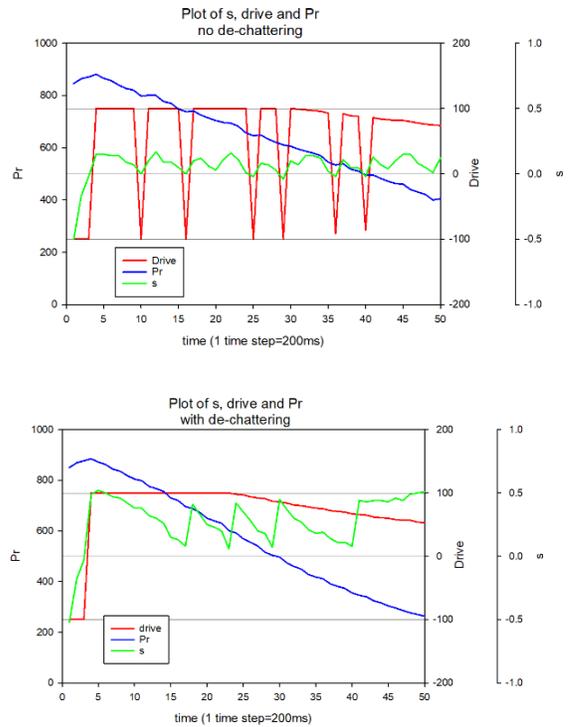

Figure 2. Comparison between non de-chattering and de-chattering

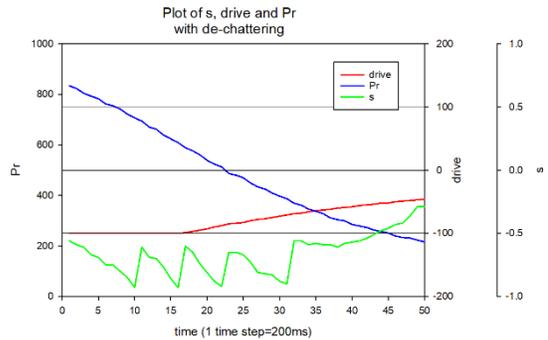

Figure 3. De-chattering from different direction

*End point detection*

The sliding mode will continue afterwards, pausing only when minimum has been reached. This minimum is

not necessary equals to zero due to mismatch. This state

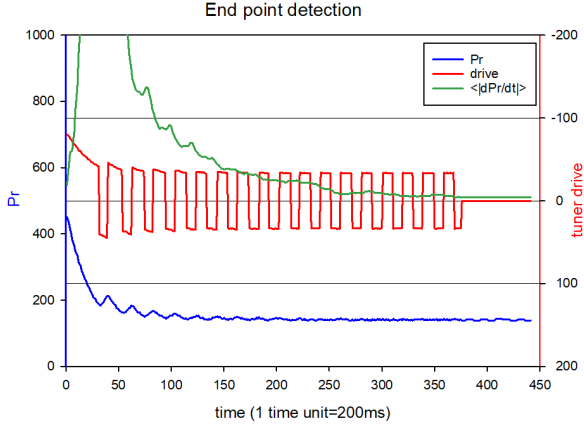

Figure 4. End point (Minimum Reflected power) detection

is indicated average of the derivative of reflected power. i.e.

$$\left\langle \left| \frac{dP_r}{dt} \right| \right\rangle < p \approx 0.1$$

A moving average of 40 samples are used for the averaging, while the differentiation is obtained from coefficients of a Savitzky-Golay filter. The results are shown in Fig.4, using a $|\Gamma| = 0.3$, where at time=150 the minimum is achieved at about $P_r = 140$. The system continues to switch directions until $\left\langle \left| \frac{dP_r}{dt} \right| \right\rangle$ becomes small enough to put the system into hibernation. As can be seen from Fig. 4, there still remains a large amount of chatters before the system finally goes into hibernation. But however this is a small amplitude tuner drive, and the effect on the reflected power is quite small. Nevertheless, in an effort to reduce this chatter, $\rho$ is reduced by a factor of $\frac{1}{2}$ when the reflected power $P_r$ has been reduced to below a preset amount closed to the minimum. But since we cannot reduce $\rho$ without affecting $s$ in Eq. 3, we reduce $\Delta t$ instead. This is illustrated in Fig. 5. The effect of the reducing of $\rho$ occurs at $t \simeq 30$, at which $P_r = 250$. As can be seen in the plot this reduction resulted in a much gentler slope in $s$. This significantly reduces the amount of chatter and therefore $\left\langle \left| \frac{dP_r}{dt} \right| \right\rangle$, allowing the system to go into hibernation much sooner. Furthermore, when the reflected power is below a threshold and $\left\langle \left| \frac{dP_r}{dt} \right| \right\rangle > 0$, indicating the power starts to rise again, the system immediately goes into hibernation. These are summarized in the following snippets to reduce $\Delta t$. The first part is called "ρ reduction", if ( reversePower < 1.25*slidingMode.deadband) dtx = 0.5*deltaTime;
if ( reversePower < 1.00*slidingMode.deadband) dtx = 0.25*deltaTime;

and to go into hibernation using the "end detection":

if (fabs(devAvg.val.avg) < 0.5 && reversePower < 1.5*slidingMode.deadband) {...}
if (devAvg.val.avg < 1.0 && reversePower < slidingMode.deadband) {...}
if (SGAvgRev.val.der>0 && reversePower < slidingMode.deadband) {...}

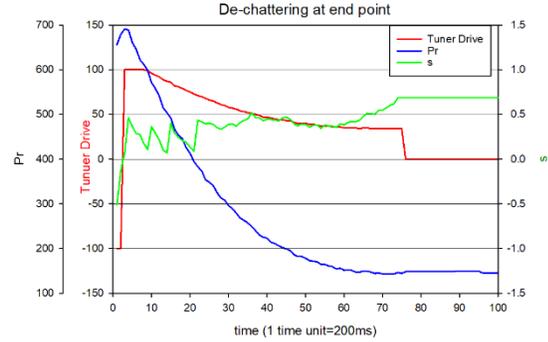

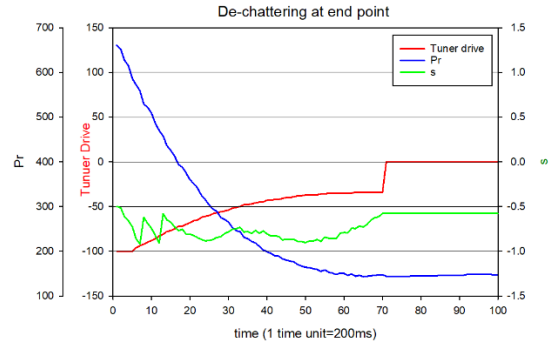

Figure 5. Dechattering at end point by reducing $\Delta t$

Summarizing the results, Fig. 6 show the comparison of responses between a non de-chattered controller and a de-chattered controller. Both systems invoke close loop control at t=20 with the same optimized $\rho = 75$. As can be seen from the figure, the non dechattered system shows a lot of chattering, starting at t=35, while the de-chattering system the switch surface is protected by the halo hardly shows any chatter. Although the convergence speeds are about the same, the de-chattered system will result in less mechanical stress and wear in the tuning mechanism. For non optimized $\rho = 150$, which is too large, the results are more distinct. This is shown in Fig. 7, where for the de-chattered system the convergence is similar to that of $\rho = 75$, but the non-chattered system there are a lot of chatters and the system take as much as twice longer to converge. If the response is too small, there will be a lot of chatter at mid-range of the reflected power for the non de-chattered case. But when the reflected power drops below 250, ρ falls on the correct range and the power reduction is smooth and has a lot less chatter. Conversely for the de-chattered case at the mid range the switching is protected by the halo and avoids chattering. However, at $P_r < 300$ the ρ-reduction kicks in and since the original ρ is already too low, the reduced ρ increases the chatter until the "end detection" algorithm kicks in and stop the movement.

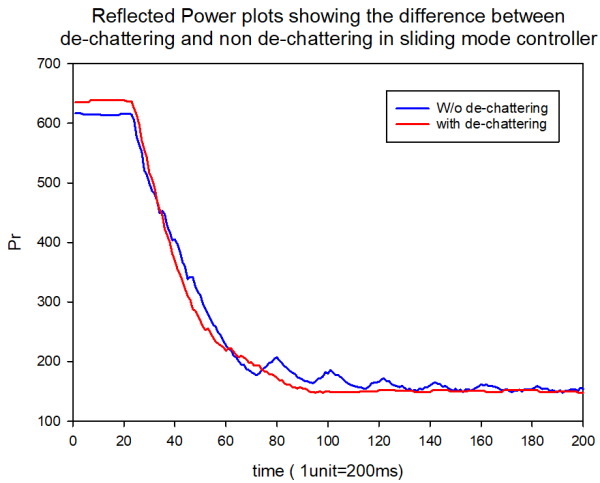

Figure 6. Comparison of responses between a non de-chattered controller and a de-chattered controller with optimized $\rho = 75$.

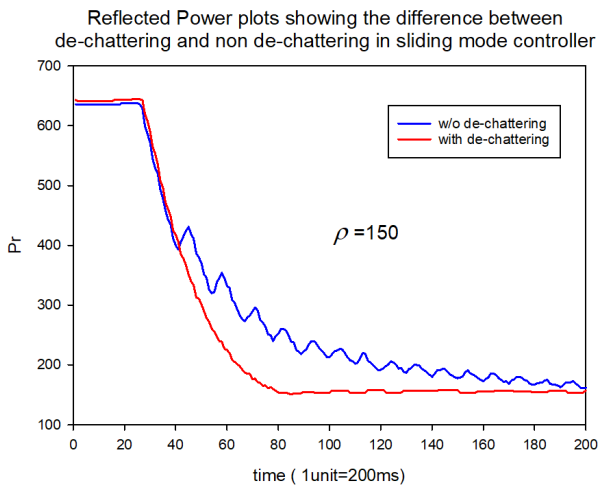

Figure 7. Comparison of responses between a non de-chattered controller and a de-chattered controller with non optimized $\rho = 150$.

## CONCLUSION

The position preset, phase alignment and sliding mode controllers will be used in the new ISAC-1 resonance control. Base on each system's strength and weakness, they will be used at different stages of powering up. The position preset is used during the initial stage of powering up, when the RF is not yet established and is still in pulse mode. When the RF level reaches a preset value, and switching from pulse to CW is successful, the control enters into phase alignment mode. At this stage the RF will continue to be ramping up. When phase alignment is completed the control will switch to sliding mode. When this average below the prefixed value the sliding mode will enter in a sleep mode, only to be awoken when this value is exceeded the threshold. When choosing $\rho$ for the sliding mode controller, it is advantageous to choose a larger $\rho$, as the amount of chattering can be suppressed by the "switch surface skipping" method.